\documentclass[reprint,amsmath,amssZb,aps,nofootinbib,longbibliography]{revtex4-2}
\usepackage{graphicx} %
\usepackage{blindtext}
\usepackage{titlesec}
\usepackage{amsmath}
\usepackage{hyperref}
\hypersetup{
    colorlinks=true,
    citecolor=blue,
    linkcolor=blue,
    filecolor=blue,      
    urlcolor=blue,
    pdftitle={Overleaf Example},
    pdfpagemode=FullScreen,
    }

\usepackage{mathptmx}
\usepackage{mathrsfs}
\DeclareMathAlphabet{\mathcal}{OMS}{cmsy}{m}{n}
\usepackage[english]{babel}

\usepackage[utf8]{inputenc}

\usepackage{scalerel}
\usepackage{tikz}
\usetikzlibrary{svg.path}

\usepackage{soul}

\usepackage{xcolor}

\definecolor{orcidlogocol}{HTML}{A6CE39}
\tikzset{
  orcidlogo/.pic={
    \fill[orcidlogocol] svg{M256,128c0,70.7-57.3,128-128,128C57.3,256,0,198.7,0,128C0,57.3,57.3,0,128,0C198.7,0,256,57.3,256,128z};
    \fill[white] svg{M86.3,186.2H70.9V79.1h15.4v48.4V186.2z}
                 svg{M108.9,79.1h41.6c39.6,0,57,28.3,57,53.6c0,27.5-21.5,53.6-56.8,53.6h-41.8V79.1z M124.3,172.4h24.5c34.9,0,42.9-26.5,42.9-39.7c0-21.5-13.7-39.7-43.7-39.7h-23.7V172.4z}
                 svg{M88.7,56.8c0,5.5-4.5,10.1-10.1,10.1c-5.6,0-10.1-4.6-10.1-10.1c0-5.6,4.5-10.1,10.1-10.1C84.2,46.7,88.7,51.3,88.7,56.8z};
  }
}

\newcommand\orcidicon[1]{\href{https://orcid.org/#1}{\mbox{\scalerel*{
\begin{tikzpicture}[yscale=-1,transform shape]
\pic{orcidlogo};
\end{tikzpicture}
}{|}}}}

\newcommand{\hide}[1]{}
\newcommand{\headline}[1]{\textcolor{black}{#1}}

\newcommand{\updatea}[1]{\textcolor{black}{#1}}

\newcommand{\dum}[1]{#1}

\definecolor{captioncolor}{rgb}{1, 0.5, 0.0} %
\definecolor{forestgreen}{rgb}{0.13, 0.55, 0.13}



\begin{document}

\title{Interpretable machine learning of magnetic transition temperature in Heusler magnets via hierarchical dependence extraction}

\author{Jean-Baptiste Mor\'ee\orcidicon{0000-0002-0710-9880}$^{1}$}
\thanks{jean-baptiste.moree@riken.jp}

\author{Ryotaro Arita$^{1,2}$\orcidicon{0000-0001-5725-072X}}

\author{Juba Bouaziz\orcidicon{0000-0003-3605-0563}$^{1,2,3}$}
\thanks{juba.bouaziz@uni-due.de}

\affiliation{\vspace{0.5em}
$^{1}$ RIKEN Center for Emergent Matter Science, Wako, Saitama 351-0198, Japan\\
$^{2}$ Department of Physics, University of Tokyo, Bunkyo-ku, Tokyo 113-0033, Japan\\
$^{3}$ Faculty of Physics, University of Duisburg-Essen, 47057 Duisburg, Germany}

\begin{abstract}
We employ interpretable
machine learning to analyze the material dependence of the magnetic transition temperature $T_c$ in ferromagnetic and ferrimagnetic Heusler compounds.
For over 200 \dum{candidate materials} with the same $F\overline{4}3m$ crystal structure but different chemical formulae and lattice constants, we consider both experimental $T_c$ and those computed via classical Monte Carlo simulations using magnetic interactions derived from \textit{ab initio} calculations.
We use the hierarchical dependence extraction (HDE) procedure [Mor\'ee and Arita, \href{https://journals.aps.org/prb/abstract/10.1103/PhysRevB.110.014502}{Phys. Rev. B \textbf{110}, 014502 (2024)}] to determine how $T_c$ depends on chemical composition and magnetic moments, from leading to higher-order effects, and use these dependencies to construct an explicit expression for $T_c$. Our results show that the HDE framework predicts $T_c$ with accuracy comparable to other machine-learning approaches such as neural network and random forest algorithms while remaining fully interpretable. $T_c$ is primarily governed by the proportions of Fe, Co, and Mn, increasing systematically with their concentration. 
These findings clarify how chemical composition and magnetic moments influence $T_c$ in collinear Heusler alloys and support the use of the HDE for computationally guided discovery of new functional materials with tailored $T_c$ values.
\end{abstract}

\maketitle

\section{Introduction}
Heusler compounds constitute a large family of ordered intermetallics, first identified in the ferromagnetic alloy $\mathrm{Cu}_2\mathrm{MnAl}$~\cite{heusler1903uber}. The half-Heusler and full-Heusler phases crystallize in the noncentrosymmetric C1$_b$ and centrosymmetric L2$_1$ cubic structures, respectively~\cite{Graf2011}. These materials exhibit a wide range of electronic and magnetic properties. Some half-metallic Heuslers show metallic behavior in one spin channel and a band gap in the other~\cite{deGroot1983}. They can also host topological electronic phases, including Weyl and Dirac semimetals, which produce large anomalous, spin, and topological Hall effects relevant for transport studies~\cite{Manna2018}. Additionally, certain Heuslers have been identified as promising thermoelectric materials~\cite{fu2015realizing}. 

Magnetically, Heusler compounds display a variety of ordered states, including ferromagnetic~\cite{deGroot1983,Jourdan2014}, ferrimagnetic~\cite{Ouardi2013SGS}, and antiferromagnetic~\cite{Maca2012CuMnAs_AFMCuMnSb} ground states. Some compounds, such as Pt$_2$MnGa and Mn$_2$RhSn, exhibit complex noncollinear and long-range helical magnetic structures~\cite{Singh2016Pt2MnGaHelicalAFM,Meshcheriakova2014Mn2RhSnNoncollinear}, while others host topological antiskyrmion spin textures~\cite{Nayak2017Antiskyrmions,Jena2019AntiskyrmionMn2RhIrSn}. The magnetic exchange interactions depend sensitively on atomic species, site occupancies, and chemical order, which allows for tunable transition temperatures between magnetically ordered and paramagnetic phases. Reported transition temperatures span a broad range, from approximately $T_c \simeq 10$~K to 1261~K~\cite{KACZMARSKA1998210,AHMAD2019599,Kubler1983,Wurmehl2006,Graf2011,Felser2015,Kundu2017}.
These magnetic properties enable a range of different applications for Heusler alloys~\cite{Quinn2021,Tavares2023,Wederni2024}. The half-metallicity and high spin polarization are exploited for spintronic devices, e.g. spin injection~\cite{Ramsteiner2008}. They are also important for magnetocaloric technologies, as Ni-Mn-based Heuslers exhibit a giant inverse magnetocaloric effect making them promising for solid-state magnetic refrigeration~\cite{krenke2005inverse}. Noncollinear and chiral Heusler magnets further enable topological spintronics, where skyrmions and antiskyrmions act as nanoscale information carriers for racetrack-type devices~\cite{Nayak2017Antiskyrmions}.

Heusler compounds have been extensively studied by \dum{first-principles} 
methods. Density functional theory (DFT) provides a microscopic description of magnetic moment formation, electronic structure, and the nature of the exchange interactions, with magnetic exchange parameters extracted via spin-spiral calculations for example. Applying this to compounds such as Co$_2$XY (X = Ti, V, Cr, Mn; Y = Al, Ga, Sn, Si) alloys or  Ni$_2$MnX (X = Ga, In, Sn, Sb) alloys leads to accurate mean-field estimates of $T_c$ compared to experiments~\cite{Kubler2007,Sasioglu2004}. The origin of the band gap and ferromagnetism in half-metallic XMnSb (X = Co, Ni, Rh, Ir, Pd, Pt) alloys was also investigated~\cite{Galanakis2002}. More recent high-throughput DFT calculations~\cite{fortunato2024high} have explored the broader A$_2$BC family 
where A, B, and C are among the 28 transition metal elements from Sc to Hg excluding technetium, and there is at least one magnetic $3d$ atom from V to Ni on either the A or B/C sites. A total of 686 compounds were inspected focusing on the structural stability and magnetic ordering. Beyond standard DFT, many-body corrections using the G$_0$W$_0$ method were applied to compounds such as Co$_2$MnSi and Co$_2$FeSi to investigate modifications of the band
gap~\cite{Meinert2012GW_Heuslers}, while dynamical mean-field theory was employed to study strong correlations in ferrimagnetic Heuslers, including FeMnSb and Ni$_{1-x}$Fe$_x$MnSb~\cite{Chioncel2006PRL_DMFT_Heusler}.
\dum{Very recently, Xiao and Tadano performed high-throughput first-principles screening of ordered ternary Heusler alloys, explicitly considering phonon stability and magnetic transition temperatures, and identified stable magnetic candidates including compensated ferrimagnets~\cite{Xiao2025}.}

The experimental and \textit{ab initio} studies discussed above are valuable for identifying specific Heusler compounds and sub-families of interest. However, obtaining a broader understanding of how $T_c$ depends on atomic species, chemical order, and lattice constant benefits from machine-learning methods, which can extract these trends from large amounts of data~\cite{Nelson2019,Belot2023}. The first step is to collect a large set of material-dependent values of the magnetic transition temperature \(T_c\), together with the magnetic phase and the chemical formula. This can be done by high-throughput calculations~\cite{Lin2015PRB_HT_TopologicalHeuslers} or by mining the literature to build databases such as JuHemd~\cite{Kovacik2022}, which compiles experimental data and DFT+Monte-Carlo results; we use JuHemd in this work. The second step is analyze the database using machine learning. A wide range of machine-learning methods\dum{~\cite{McCulloch1943,Rosenblatt1958,Rumelhart1986,LeCun1998,Krizhevsky2012,Goodfellow2016,Schmidt2009,Udrescu2020,Cover1967,Breiman2001,Hoerl1970,Saunders1998}} has been applied to the data-driven discovery of new materials and the mining of trends in material properties~\cite{Burlacu2023,Carbone2020,Lee2023,Vu2015,Nelson2019,Belot2023,Hilgers2025,Jung2024}. Such approaches have been used to predict the magnetic transition temperature \(T_c\) in ferromagnets~\cite{Nelson2019,Belot2023,Jung2024} and in Heusler compounds~\cite{Liu2025,Hilgers2025}. Given the property we wish to predict (the target \(y=T_c\)) and a set of experimentally controllable descriptors \(\mathbf{x}=\{x_i\}\), machine learning builds a model \(y=f(\mathbf{x})\) that maps descriptor values to \(T_c\), and can predict \(T_c\) for descriptor combinations not present in the database.

\begin{figure}[!htb]
    \centering    \includegraphics[width=1\linewidth]{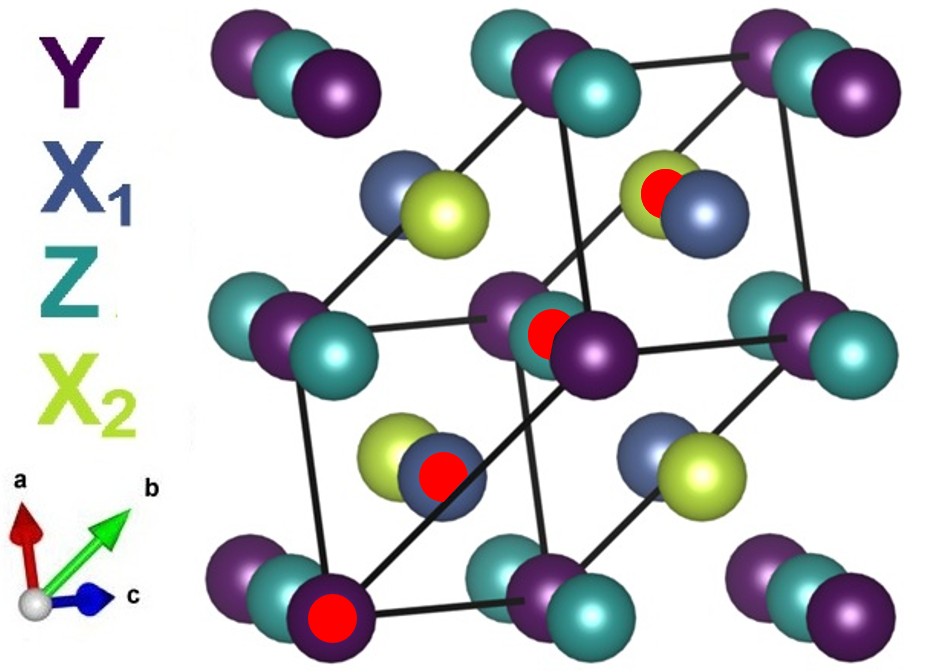}
    \caption{Crystal structure of Heusler compounds.
    The general chemical formula is denoted as X$_1$X$_2$YZ, where X$_1$, X$_2$, Y, and Z are either magnetic transition metal elements with an open $3d$ shell (from Ti to Ni) or other chemical elements included in the JuHemd database, namely $3d$ (Sc, Cu, Zn), $4d$ (Zr, Nb, Ru, Rh, Pd, Ag), and $5d$ (Hf, Ta, W, Ir, Pt, Au) transition metals, together with Be and $p$-block elements from groups 13-15 (B, Al, Ga, In, Tl, Si, Ge, Sn, Pb, As, and Sb).
    The space group is $F\overline{4}3m$ (No. 216).
    The unit cell is delimited by the black lines, and 
    the Wyckoff positions of Y, X$_1$, Z, and X$_2$ are $(0,0,0)$, $(1/4,1/4,1/4)$, $(1/2,1/2,1/2)$, and $(3/4,3/4,3/4)$, respectively.
    Atoms within the unit cell are marked by red spheres.
    We show the primitive vectors that span the unit cell, ${\bf a} = [0 \ a/2 \ a/2]$, ${\bf b} = [a/2 \ 0 \ a/2]$, and ${\bf c} = [a/2 \ a/2 \ 0]$ in Cartesian coordinates,
    where $a$ is the lattice constant.}
    \label{fig:cryst}
\end{figure}

The analysis of $T_c$ is facilitated by similarities between Heusler compounds. Despite their diverse chemical compositions, they share the same general crystal structure (see Fig.~\ref{fig:cryst}), which makes them a convenient platform for methods that seek to relate materials composition to $T_c$. However, the machine-learning procedures previously used to extract the material dependence of $T_c$ are not interpretable by humans. To gain more physical insights, a procedure must both predict $T_c$ accurately and reveal how $T_c$ depends on material variables. Accuracy is commonly quantified by the coefficient of determination $R^2$; previous studies on ferromagnets and Heusler magnets report $R^2$ values in the range $0.53$ to $0.93$ (see Appendix~\ref{app:mlworks})~\cite{Nelson2019,Belot2023,Hilgers2025,Jung2024}. Those works also identified clear chemical drivers of $T_c$, notably Fe and Co content and the number of $d$ valence electrons. On the one hand, methods that achieve the best predictive performance are often implicit, black-box models that do not provide an explicit, physically meaningful expression for $T_c$.  On the other hand, explicit models can return closed-form expressions, but these are frequently overly complex or depend on many descriptors.
This prevents a clear, quantitative ranking of which variables control $T_c$, which is essential to discover and synthesize new magnetic materials that function at different temperatures.

In this paper we employ the hierarchical dependence extraction (HDE) procedure~\cite{Moree2024HDE}. HDE is a supervised symbolic-regression method that constructs an interpretable model iteratively by adding monotonic, single-descriptor functions ranked by importance. Simple power-law forms keep the model easy to interpret and comparable with physical scaling laws, while the structured search makes HDE computationally cheaper than brute-force searches~\cite{Udrescu2020} or evolutionary algorithms~\cite{Cranmer2023}. We analyse $T_c$ as a function of descriptors such as chemical proportions and magnetic moment amplitudes stored in the JuHemd database~\cite{Kovacik2022}. We restrict our study to collinear ferromagnetic (FM) and ferrimagnetic (FiM) Heusler alloys, for which the magnetic moments can be treated as scalar quantities (along the $z$ direction). These materials exhibit a wide distribution of $T_c$ values~\cite{Kundu2017}, reaching $T_c \simeq 1100$ K for Co$_2$FeSi~\cite{Wurmehl2006} with corresponding variations in magnetic moments~\cite{Kovacik2022}. Using this approach, we extract the quantitative, hierarchical dependence of $T_c$ on chemical proportions (experimentally controllable) and on \textit{ab initio} element-resolved magnetic moment amplitudes (intermediate quantities). This choice of intermediate quantities is motivated by explicit expressions for $T_c$ that involve the magnetic moments, derived from an analytical expansion of the magnetic exchange interactions in combination with mean-field theory~\cite{Yoshida1996_TheoryOfMagnetism} or the random-phase approximation (RPA) to obtain $T_c$~\cite{Turek2006}.

{This paper is organized as follows.}
Section~\ref{sec:framework}
describes the database, the choice of descriptors, and a
reminder of the HDE approach. Section~\ref{sec:results} gives the HDE results and the hierarchical dependence of $T_c$. Section~\ref{sec:disc} discusses the comparison of the HDE with previously published machine learning approaches, and the physical insights within our HDE approach. Appendix~\ref{app:mlworks} summarizes previous machine learning works on magnetic materials including Heusler alloys. Appendix~\ref{app:mfrpa} gives the analytical expressions of $T_c$ derived within the mean-field and RPA approaches.

\section{Methodology}
\label{sec:framework}
We describe the methodology employed to extract the general dependence of $T_c$ in Heusler compounds with collinear magnetic order. We start from the JuHemd database,
then select the target $y=T_c$ and build descriptors ${\bf x} = \{ x_i \}$,
then apply the HDE to obtain the dependence of $T_c$. The three steps involved in this framework are explained in detail below.

\subsection{Database}
\label{sec:meth-data}

The JuHemd database~\cite{Kovacik2022} contains material-dependent physical quantities for a collection of $N_c = 400$ Heusler compounds. It includes both experimentally measured quantities and \textit{ab initio} results obtained using DFT together with the local density approximation (LDA) and the generalized gradient approximation (GGA) for the exchange-correlation energy. In this paper, we focus on the experimental quantities and on the GGA data (the latter were also examined in Ref.~\cite{Hilgers2025}).  For the experimental dataset, $N_c = 186$ compounds exhibit FM or FiM order, giving $N_{\rm ent} = 653$ entries. For the calculated dataset, $N_c = 206$ compounds exhibit FM or FiM order, giving $N_{\rm ent} = 372$ entries.\footnote{The number of entries for compounds that are not FM or FiM is only $\simeq 36$, so that FM and FiM compounds constitute the vast majority of the database.} For a given compound, multiple entries correspond to different experiments or calculations. The chemical formula is identical across entries, but the atomic arrangement and/or experimental conditions may differ, leading to different values of $T_c$ and intermediate quantities.

For a given entry, the key available quantities are summarized below. Each entry includes the value of $T_c$ (denoted as $T_c^{\rm exp}$ for an experimental entry and $T_c^{\rm calc}$ for a GGA entry), the magnetic ordering at temperatures $T < T_c$, and the lattice constant $a$. Calculated entries also include the element and magnetic moment on each site $i = 1,2,3,4$ in the unit cell (respectively the red, blue, green and yellow \dum{atoms marked by red spheres} in Fig.~\ref{fig:cryst}). For each site $i$ there are $N(i)$ different configurations representing chemical disorder; $N(i)=1$ if the site is chemically ordered. For each configuration $j = 1,\dots,N(i)$ the entry includes $(\mathrm{Y}_{i,j}, c_{i,j}, \mu_{\mathrm{Y}, i, j})$, where $\mathrm{Y}_{i,j}$ is the chemical element on site $i$ in configuration $j$, $c_{i,j}$ is its concentration on site $i$ in that configuration (so $\sum_{j=1}^{N(i)} c_{i,j}=1$), and $\mu_{\mathrm{Y}, i, j}$ is the magnetic moment of element $\mathrm{Y}_{i,j}$ on site $i$ in configuration $j$.

The computational approach used to build the JuHemd database is summarized here. The electronic structure calculations rely on the full potential all-electron Korringa-Kohn-Rostoker Green function method~\cite{jukkr3.6} using the GGA~\cite{Perdew1992} for the exchange-correlation energy. The mapping of the \textit{ab initio} Hamiltonian onto a classical Heisenberg model with localized spins is done using the infinitesimal rotation method~\cite{Liechtenstein1987,Szilva2023}. The spin model Hamiltonian is then solved using Monte-Carlo simulations~\cite{Metropolis1953} to evaluate $T_c^{\rm calc}$. We note that the Heisenberg Hamiltonian assumes that longitudinal fluctuations of the magnetic moments can be neglected, i.e., the magnetic moments are local and robust. This assumption will be supported by the quantitative agreement between the HDE expressions for $T_c^{\rm exp}$ and $T_c^{\rm calc}$, as discussed later.

\subsection{Choice of target and descriptors}
\label{sec:meth-var}

\begin{figure}[!htb]
    \centering
    \includegraphics[width=0.95\linewidth]{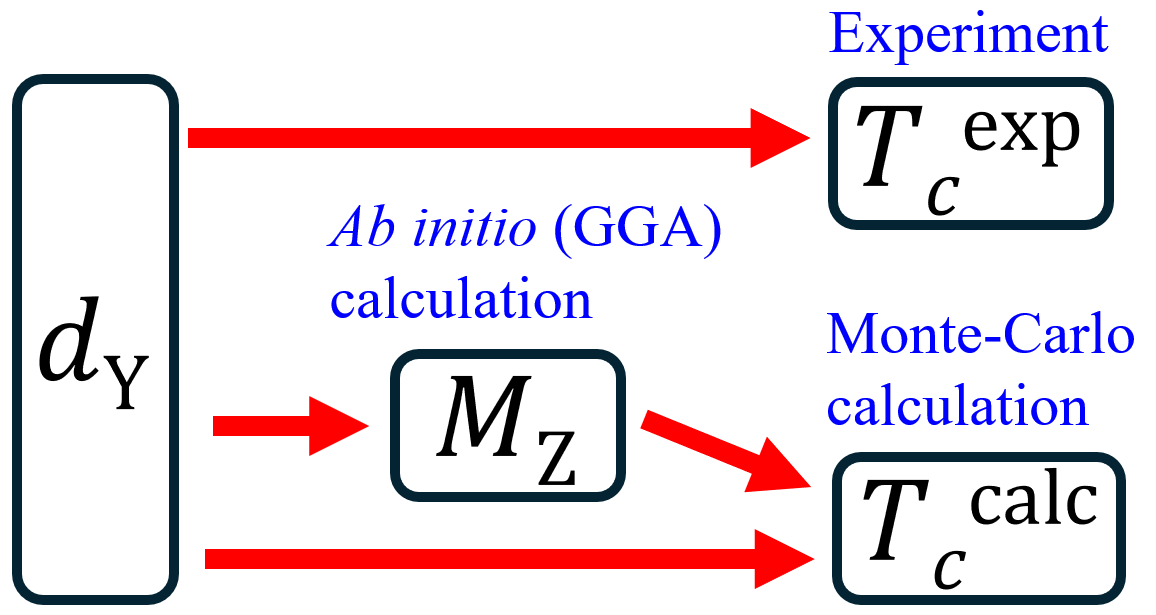}
\caption{Schematic representation of the dependencies analyzed in this work. 
The analysis begins with the chemical proportions $d_{\rm Y}$ and 
the products $d_{\rm Y} d_{\rm Z}$, which follow directly from the 
chemical formula;
\dum{Y is any chemical element included in the JuHemd database, while Z is restricted to $3d$ transition 
metals between Ti and Ni.} 
First, the dependence of the experimental 
and calculated magnetic transition temperatures, $T_c^{\rm exp}$ and 
$T_c^{\rm calc}$, on $d_{\rm Y}$ and $d_{\rm Y} d_{\rm Z}$ is examined. 
Second, the dependence of $T_c^{\rm calc}$ on the element-resolved 
magnetization amplitudes $M_{\rm Z}$ and their products $M_{\rm Z} M_{\rm Z'}$, 
obtained by postprocessing the \textit{ab initio} results (Z and Z$'$ 
are magnetic elements between Ti and Ni), is analyzed. Lastly, the 
dependence of the magnetization amplitudes $M_{\rm Z}$ on $d_{\rm Y}$ 
and $d_{\rm Z} d_{\rm Y}$ is investigated.}
\label{fig:hdevar}
\end{figure}

The choice of target and descriptors for the HDE procedure 
is illustrated in Fig.~\ref{fig:hdevar}. The target $y$ is the calculated 
(experimental) magnetic transition temperature  $T_c^{\rm calc}$ 
($T_c^{\rm exp}$). The choice of descriptors ${\bf x} = \{x_i\}$ is 
motivated by the preference for physical quantities that can be controlled 
experimentally. The first elemental quantity to consider is the 
the chemical proportion $d_{\rm Y}$ of the element Y in a given compound, this is controllable experimentally by tuning the chemical formula. The value of $d_{\rm Y}$ is between 0 and 1 (0 if Y is absent from the compound, 1 if the compound includes only Y). For Heusler compounds, the possible values of $d_{\rm Y}$ are 0, 0.25, 0.50, and 0.75, because there are $N_{\rm sites} = 4$ atoms in the unit cell shown in Fig.~\ref{fig:cryst}, and the number of Y atoms in the unit cell can be 0, 1, 2, or 3. The value of $d_{\rm Y}$ is straightforwardly deduced from the chemical formula; it is also possible to express $d_{\rm Y}$ as
\begin{equation}
    d_{\rm Y} = \frac{1}{N_{\rm sites}} \sum_{i}^{N_{\rm sites}} \sum_{j=1}^{N(i)} \delta_{{\rm Y},{\rm Y}_{i,j}} c_{i,j} ,
\end{equation}
where $c_{i,j}$ is the chemical concentration on site $i$ in the configuration $j$,
and $\delta_{{\rm Y},{\rm Y}_{i,j}}$ is 1 if Y = Y$_{i,j}$ and 0 if Y $\neq$ Y$_{i,j}$.

The second intuitive approach is to construct an intermediate quantity based on the \textit{ab initio} element-resolved magnetization amplitudes in order to identify which chemical elements primarily control $T_c$. To do this, we decompose the total magnetization $M$ into its elemental contributions and use the element-resolved magnetization amplitude $M_{\rm Z}$ as a descriptor of the element-specific influence on $T_c$. We therefore define $M_{\rm Z}$ as:
\begin{equation}
    M_{\rm Z} = \frac{1}{V_{\rm cell}} \sum_{i}^{N_{\rm sites}} \sum_{j}^{N(i)} \delta_{{\rm Z},{\rm Z}_{i,j}} c_{i,j} |\mu_{{\rm Z}, i, j}|
    \label{eq:my}
\end{equation}
where
Z is a magnetic element with an open $3d$ shell from Ti to Ni (Z forms the subset of Y that is relevant to magnetism),
and
$V_{\rm cell} = a^3/4$ is the volume of the unit cell generated by the primitive vectors in Fig.~\ref{fig:cryst}. $M_{\rm Z}$ is indirectly controllable by tuning the chemical formula, since $|\mu_{{\rm Z}, i, j}|$ implicitly depends on the chemical concentration $c_{i,j}$ and element ${\rm Z}_{i,j}$ on site $i$ in the configuration $j$. However, we do not have an explicit expression of $|\mu_{{\rm Z}, i, j}|$; instead we rely on HDE to express $M_{\rm Z}$ as a function of the chemical proportions. Our choice of $M_{\rm Z}$ as intermediate quantities is motivated by guidance from mean-field theory and RPA. Indeed, in Appendix~\ref{app:mfrpa}, we derive explicit mean-field and RPA expressions of $T_c$ [see Eqs.~\eqref{eq:tcmf} and~\eqref{eq:tcrpa}], and we find that $T_c$ scales with the magnetic moment amplitude, and vanishes when the latter vanishes. In addition, in the Landau theory of ferromagnets, the magnetization along the chosen axis, here $z$, can be expressed in terms of the temperature $T$ and $T_c$, vanishing at $T = T_c$~\cite{coey2010magnetism}.

{Note that we restrict $M_{\rm Z}$ to the subset Z (the elements with open $3d$ shell from Ni to Ti), which is the magnetically active subset in Heusler compounds.
Small induced magnetic moments whose amplitude is typically $\lesssim 0.1 \mu_{\rm B}$ may appear on non-Z sites, but this is due to their spatial proximity with magnetic Z sites, and the induced magnetic moments are not independent from the Z subset (see e.g. Ref.~\cite{Sasioglu2008} with Z = Mn).
Thus, we do not take into account the magnetization amplitude for non-Z elements.
The origin of the small induced magnetic moments is the strong intra-atomic exchange splitting
of Z $d$ states in specific compounds, e.g. in \dum{X$_2$ZY} with Z = Mn, X = Co, Ni, Cu, Pd, and Y = Al, Sn, In, Sb~\cite{Kubler1983}: The exchange-splitted Z $d$ states hybridize with neighboring atoms, which may induce the small magnetic moments.} {Moreover, even within the Z subset, magnetic moments can be either intrinsic or induced depending on which atoms surround the Z site.
While certain Z sites may host robust magnetic moments whose amplitude is more than $1 \mu_{\rm B}$, other Z sites carry only small induced magnetic moments whose amplitude is $\lesssim 0.1-0.5 \mu_{\rm B}$.
Thus, only moments whose amplitude exceeds a given threshold $\mu_{\rm thr}$ are included in $M_{\rm Z}$, while smaller moments are considered induced and are neglected in $M_{\rm Z}$.
This is achieved by setting $|\mu_{{\rm Z}, i, j}| = 0$ in Eq.~\eqref{eq:my} if $|\mu_{{\rm Z}, i, j}| < \mu_{\rm thr}$. In this work, two threshold moments are considered: $\mu_{\rm thr} = 0\mu_{\rm B}$ and $\mu_{\rm thr} = 0.5\mu_{\rm B}$.
As seen later, the HDE result has little dependence on the choice of $\mu_{\rm thr}$.}

Lastly, in the practical HDE scheme, we include not only the chemical proportions and magnetization 
amplitudes but also their pairwise products, which carry additional information. Specifically, 
we use ${\bf x} = \{ d_{\rm Y}, d_{\rm Y} d_{\rm Z} \}$ instead of ${\bf x} = \{ d_{\rm Y} \}$,
and ${\bf x} = \{ M_{\rm Z}, M_{\rm Z} M_{\rm Z'} \}$ instead of ${\bf x} = \{ M_{\rm Z} \}$,
where Z and Z' denote magnetic elements with an open $3d$ shell (Ti--Ni), and Y is any 
chemical element \dum{included in the JuHemd database}.

\subsection{Hierarchical dependence extraction}
\label{sec:meth-hde}

\headline{Here, we summarize the essence of the HDE procedure, whose details can be found in Ref.~\cite{Moree2024HDE}.}
We start from a given target $y$ and descriptors ${\bf x} = \{ x_i \}$.
We denote the procedure as HDE$[y,{\bf x}]$.
This procedure constructs the HDE descriptor $z$ which captures the nonlinear dependence of $y$ on the $x_i$,
then the HDE expression $y^{\rm HDE}$ which is an affine function of $z$ and gives an explicit approximation of $y$.
All these quantities have $N_{\rm ent}$ values that correspond to the $N_{\rm ent}$ entries in the data set.
\\

The HDE descriptor $z$ is constructed iteratively by adding monotonic, single-descriptor functions ranked by performance. 
At generation $g = 1$,
the expression of $z$ is
\begin{equation}
    z_{(1)}  =  x_{i_1}^{\alpha_1}, \label{eq:x1}
\end{equation}
and the variational parameters $i_1$ and $\alpha_1$ are optimized to maximize the fitness function
$|\rho[y,z^{}_{(1)}]|$,
which is the absolute value of the Pearson correlation coefficient $\rho$ between $y$ and $z_{(1)}$.
The optimized values $i_1^{\rm opt}$ and $\alpha_1^{\rm opt}$ give the HDE descriptor $z^{\rm opt}_{(1)}$. 
The descriptor $x_{i_1^{\rm opt}}$ is identified as the most important descriptor for $y$.
At generation $g \geq 2$,
the expression of $z$ is 
\begin{equation}
    z_{(g)}  =  z_{(g-1)}^{\rm opt} \Bigg[ 1 + \zeta_{g}\frac{x_{i_g}^{\alpha_g}}{[z_{(g-1)}^{\rm opt}]^{\beta_g} } \Bigg], \label{eq:xg}
\end{equation}
and the variational parameters $i_g$, $\alpha_g$, $\zeta_g$, and $\beta_g$ are optimized to maximize $|\rho[y,z^{}_{(g)}]|$
and obtain $z^{\rm opt}_{(g)}$.
The descriptor $x_{i_g^{\rm opt}}$ is identified as the $g^{\rm th}$ most important descriptor for $y$.
Details on the optimization procedure are given in Ref.~\cite{Moree2024HDE}.
The value of $\beta_g$ is restricted to be either 0 or 1 as in Ref.~\cite{Moree2024HDE}, and we impose $\alpha_g > 0$ in this paper to avoid divergences when the value of the chemical descriptor $x_{i_g}$ becomes zero, e.g. when the proportion of a given chemical element is absent. After obtaining $z^{\rm opt}_{(g)}$, we construct the HDE expression at generation $g$ as
\begin{equation}
y^{\rm HDE}_{(g)} = a_{(g)} + b_{(g)} z^{\rm opt}_{(g)},
\label{eq:ff}
\end{equation}
where the offset coefficient $a_{(g)}$ and the slope coefficient $b_{(g)}$ are obtained by an affine regression of $y$ on $z^{\rm opt}_{(g)}$.
We denote the coefficient of determination at generation $g$ as
\begin{equation}
    R^{2}_{(g)} = |\rho[y,y^{\rm HDE}_{(g)}]|^2,
    \label{eq:R2}
\end{equation}
which is also equal to $|\rho[y,z^{\rm opt}_{(g)}]|^2$.
Eq.~\eqref{eq:R2} allows us to compare the accuracy of HDE and other machine learning procedures, because the latter employ $R^2$ as well.
However, they treat $R^2$ as a post-optimization metric,
while HDE treats $R^2$ as the fitness function, because optimizing $|\rho[y,z^{}_{(g)}]|$ is equivalent to optimizing $R^2$.
We choose $R^2$ as the fitness function because its value does not depend on the scale of $y$ and $x_i$ and is thus easier to interpret than a scale-dependent fitness function.
(For instance, $R^2_{(g)} \simeq 0.8-0.9$ usually means the description of $y$ is reasonably accurate at generation $g$.)
For completeness, we also compute and show the mean average error (MAE) at generation $g$,
\begin{equation}
    {\rm MAE}_{(g)} = \frac{1}{N_{\rm ent}} \sum_{e=1}^{N_{\rm ent}} |y[e] - y^{\rm HDE}_{(g)}[e]|,
\end{equation}
where $y[e]$ denotes the value of $y$ for the entry $e$.

We stop the iteration of $g$ when the value of $R^{2}_{(g)}$ is converged.
This value is denoted as $R^{2}_{\infty}$ and is the one we compare to $R^2$ obtained in other machine learning procedures.
In practice, $R^{2}_{\infty}$ is reached at $g \updatea{\ \lesssim 10-30}$, so that we restrict to this range of $g$.
The dependence of $y$ is mainly contained in $y^{\rm HDE}_{(g)}$ at low $g \lesssim 3$, which we analyze in this paper.
Note that the risk of overfitting in the HDE procedure at $g \leq 30$ is mitigated by the reasonably high ratio between the number of entries and the number of parameters.
In the polynomial-like HDE expression, the number of $\alpha$ and $\zeta$ parameters is $N_{\rm par} = 2g-1 \leq 59$, so that $N_{\rm ent}/N_{\rm par} \geq 6.3$ for calculated entries and $N_{\rm ent}/N_{\rm par} \geq 11.1$ for experimental entries. Overfitting typically happens when $N_{\rm ent}/N_{\rm par} \sim 1$, which is not the case here.
\\

To support the reliability of the HDE result, we perform two distinct procedures. First, 
we examine if the prediction of $x_{i_g^{\rm opt}}$ for $g \leq 3$ is robust, by performing 
a score analysis as done in Ref.~\cite{Moree2024HDE}. At fixed $g$, the score of a given 
descriptor is defined as 
\begin{equation}
    S_{(g)}(x_i) = |\rho[y,z_{(g)}^{i}]|^2/|\rho[y,z_{(g)}^{\rm opt}]|^2,
\end{equation}
where $z_{(g)}^{i}$ is the HDE descriptor that maximizes the fitness function under the 
constraint $x_{i_g} = x_i$. Namely, $z_{(g)}^{i}$ is calculated by optimizing all variational 
parameters except $i_g$, whereas $z_{(g)}^{\rm opt}$ is calculated by optimizing all variational 
parameters including $i_g$. For the optimal $x_{i_g^{\rm opt}}$, the score is always 1.
For other descriptors, the score is lower than 1. If there is no other descriptor whose score 
is close to 1, the prediction is deemed robust.

Second, we examine if the expression of $y^{\rm HDE}_{(g)}$ for $g \leq 3$ is robust, 
by performing the $4$-fold cross-validation. At fixed $g$, this is done as follows.
The $N_{\rm ent}$ entries are randomly shuffled and divided into $k=4$ folds. For $l$ 
from $1$ to $k$, we perform HDE[$y$,$\bf{x}$] using the training set (all the folds except 
$l$), and obtain the HDE expression $y^{{\rm HDE}}_{(g)}$. We then use this expression 
to predict the restriction $y^{(l)}$ of $y$ to the test set (the fold $l$). The prediction 
is denoted as $y^{{\rm HDE}(l)}_{(g)}$. After the loop on $l$, we calculate $R^2$ as:
\begin{equation}
R^2_{(g)\mathrm{CV}} \;=\; 1 \;-\; 
\frac{
    \displaystyle \sum_{l=1}^{k} \sum_{e=1}^{N_{{\rm ent},l}} \left( y^{(l)}_{}[e] - y_{(g)}^{{\rm HDE}(l)}[e] \right)^2
}{
    \displaystyle \sum_{l=1}^{k} \sum_{e=1}^{N_{{\rm ent},l}} \left( y^{(l)}_{}[e] - \bar{y} \right)^2
},
\label{eq:R2cv}
\end{equation}
where $N_{{\rm ent},l}$ denotes the number of entries in the fold $l$, which verifies
\begin{equation}
    N_{\rm ent} = \sum_{l=1}^{k} N_{{\rm ent},l},
\end{equation}
and
\begin{equation}
\bar{y} \;=\; \frac{1}{N_{\rm ent}} \sum_{l=1}^{k} \sum_{e=1}^{N_{{\rm ent},l}} y^{(l)}[e]
\end{equation}
is the mean value of $y$ over all folds. Note that Eqs.~\eqref{eq:R2} and~\eqref{eq:R2cv} are equivalent if $k=1$, but not if $k>1$ in the general case.

\section{Hierarchical dependence of $T_c$}
\label{sec:results}

\subsection{Dependence of $T_c$ on chemical proportions}
\label{sec:results12}

We now use the HDE procedure explained in Sec.~\ref{sec:meth-hde} to express $T_c$ as a function of the chemical composition. We perform two independent HDE runs, one using the experimental entries and the other using the computed GGA entries. The corresponding target variables are $T_{c}^{\rm exp}$ and $T_{c}^{\rm calc}$, respectively. The descriptor set
${\bf x} = \{ d_{\rm Y},\, d_{\rm Z} d_{\rm Y} \}$ consists of the chemical proportions $d_{\rm Y}$ and their pairwise products $d_{\rm Z} d_{\rm Y}$.
Here, Y denotes any chemical element included in the JuHemd database, namely selected $3d$ (Sc--Zn), $4d$ (Zr, Nb, Ru, Rh, Pd, Ag), and $5d$ (Hf, Ta, W, Ir, Pt, Au) transition metals, together with Be and $p$-block elements from groups 13-15 (B, Al, Ga, In, Tl, Si, Ge, Sn, Pb, As, and Sb).
Furthermore, Z denotes the subset of $3d$ transition metals with partially filled $3d$ shells (Ti--Ni).

Our first result is that $T_c$ depends primarily on the Fe, Co, and Mn fractions, and increases with increasing concentration of these elements. For $g=3$ (third generation), the HDE expressions for both $T_c^{\rm exp}$ and $T_c^{\rm calc}$ are given by:
\begin{align}
(T_c)^{\rm HDE}_{(3)}
&= a_0 + a_1
\left[
d_{\rm Co,Fe}^{\alpha_1}
+
a_2\,
d_{\rm Co,Mn}^{\alpha_2}
+
a_3\,
d_{\rm Fe}^{\alpha_3}
\right], 
\label{eq:Tc_chemicalproportions}
\end{align}
with the corresponding coefficients listed in Table~\ref{tab:TcHDEcoeff}.

\begin{figure}[!htb]
    \centering    
    \includegraphics[width=1.0\linewidth]{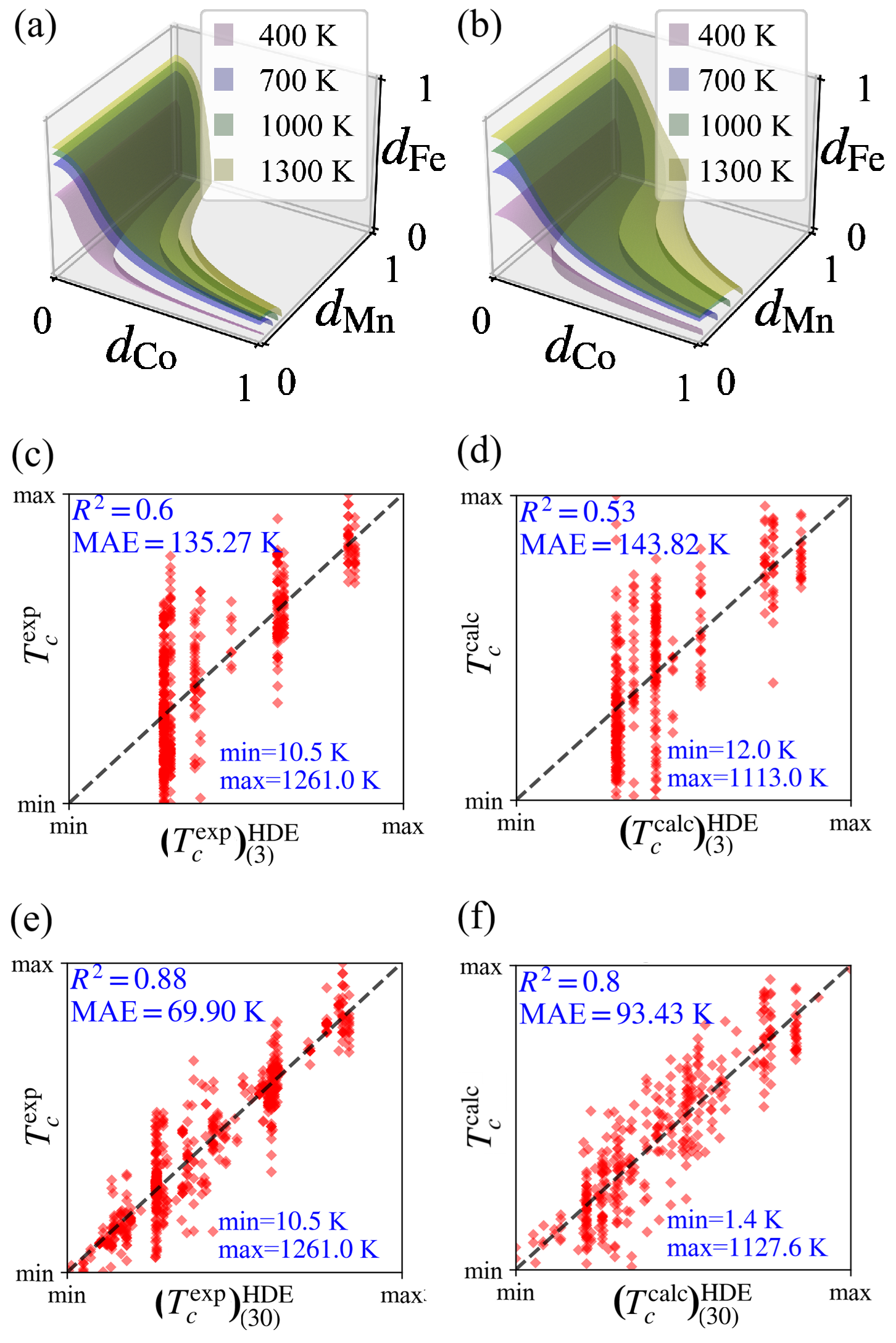}
    \caption{Results of the two independent HDE runs using the experimental entries [left column, panels (a), (c), (e)] and the calculated entries [right column, panels (b), (d), (f)].
    The target variables are respectively $T_c^{\exp}$ and $T_c^{\rm calc}$, and the descriptors are the chemical proportions of elements and their pairwise products.
    The panels (a, b) show the isosurfaces of the HDE expressions of $T_c^{\exp}$ and $T_c^{\rm calc}$ at generation 3 as a functions of the chemical proportions $d_{\rm Fe}$, $d_{\rm Co}$, and $d_{\rm Mn}$ of Fe, Co, and Mn.
    The other panels show the comparison between actual and HDE-predicted values of $T_c^{\exp}$ and $T_c^{\rm calc}$ at generation 3 [panels (c, d)] and at generation 30 [panels (e, f)].
    The $g=3$ expressions capture the dominant physical trends, while the $g=30$ expressions are the converged HDE predictions.
    The values of the coefficient of determination $R^2$ and the mean average error (MAE) are also given.}
    \label{fig:hde_Tcfy}
\end{figure}

\begin{table}[t]
\caption{Obtained coefficients $\{ a_i \}$ and exponents $\{ \alpha_i \}$ for the third-generation ($g=3$) HDE expressions of $T_c^{\rm exp}$ and $T_c^{\rm calc}$ as functions of the chemical proportions and their pairwise products. The coefficients $a_0$ and $a_1$ are in K while $a_2$ and $a_3$ are dimensionless.
We also list the values of the coefficient of determination $R^2$ at the third generation.}
\label{tab:TcHDEcoeff}
\begin{ruledtabular}
\begin{tabular}{c cc}
 & $T_c^{\rm exp}$ & $T_c^{\rm calc}$ \\
\hline
$a_0$ & 367.50 & 341.06 \\
$a_1$ & 89159.02 & 571733.22 \\
$a_2$ & 0.242 & 0.00489 \\
$a_3$ & 0.0401 & 0.00239 \\
$\alpha_1$ & 2.34 & 3.41 \\
$\alpha_2$ & 1.89 & 1.11 \\
$\alpha_3$ & 7.23 & 3.38 \\
$R^2$      & 0.60 & 0.53 \\
\end{tabular}
\end{ruledtabular}
\end{table}
The coefficients of determination are $R^2_{(3)} = 0.60$ for $T_c^{\rm exp}$ and $R^2_{(3)} = 0.53$ for $T_c^{\rm calc}$, respectively. These values indicate that the $g=3$ HDE expressions provide only a qualitative description of $T_c$ as seen in Fig.~\ref{fig:hde_Tcfy}(c,d). Nevertheless, they correctly capture the dominant physical mechanisms determining $T_c$. Moreover, the two HDE expressions at $g=3$ share the same functional form and differ only in the fitted coefficients obtained from the experimental and GGA datasets, respectively. This indicates that both models encode the same underlying descriptor structure, with different parameter values. The corresponding isosurfaces as functions of $d_{\rm Fe}$, $d_{\rm Co}$, and $d_{\rm Mn}$ show a comparable structure as seen in Fig.~\ref{fig:hde_Tcfy}(a,b). This consistency indicates that the computational DFT+Monte Carlo procedure provides a stable and physically meaningful estimate of $T_c$.

The robustness of the HDE expressions is supported by the score analysis and the $4$-fold cross-validation. In the score analysis, the second-largest score values remain well below $1$, indicating that no alternative descriptor competes closely with the optimal one. For $T_c^{\rm exp}$, the second-largest scores are $0.56$, $0.74$, and $0.86$ at $g=1$, $2$, and $3$, respectively. The corresponding values for $T_c^{\rm calc}$ are $0.88$, $0.67$, and $0.90$. In the $4$-fold cross-validation, we obtain $R^2_{(3)\mathrm{CV}} = 0.59$ for $T_c^{\rm exp}$ and $R^2_{(3)\mathrm{CV}} = 0.51$ for $T_c^{\rm calc}$, close to the corresponding training values $R^2_{(3)} = 0.60$ and $R^2_{(3)} = 0.53$.

However, the description of $T_c$ solely in terms of the chemical proportions remains limited. Indeed, the offset $a_{(g)}$ in the HDE expression remains finite even in the limit $g \rightarrow \infty$. For example, in the HDE expression for $T_c^{\rm calc}$, we obtain $a_{(3)} \simeq 341$ K and $a_{(30)} \simeq 241$ K. Consequently, $T_c^{\rm calc}$ does not vanish when all the magnetic chemical proportions are zero, i.e., when $z^{\rm opt}_{(g)} = 0$. This limitation is addressed in the next subsection by expressing $T_c^{\rm calc}$ in terms of the element-resolved magnetization amplitudes.

\subsection{Dependence of calculated $T_c$ on element-resolved magnetization amplitudes}
\label{sec:results3}
\begin{figure}[!htb]
    \centering     \includegraphics[width=1.0\linewidth]{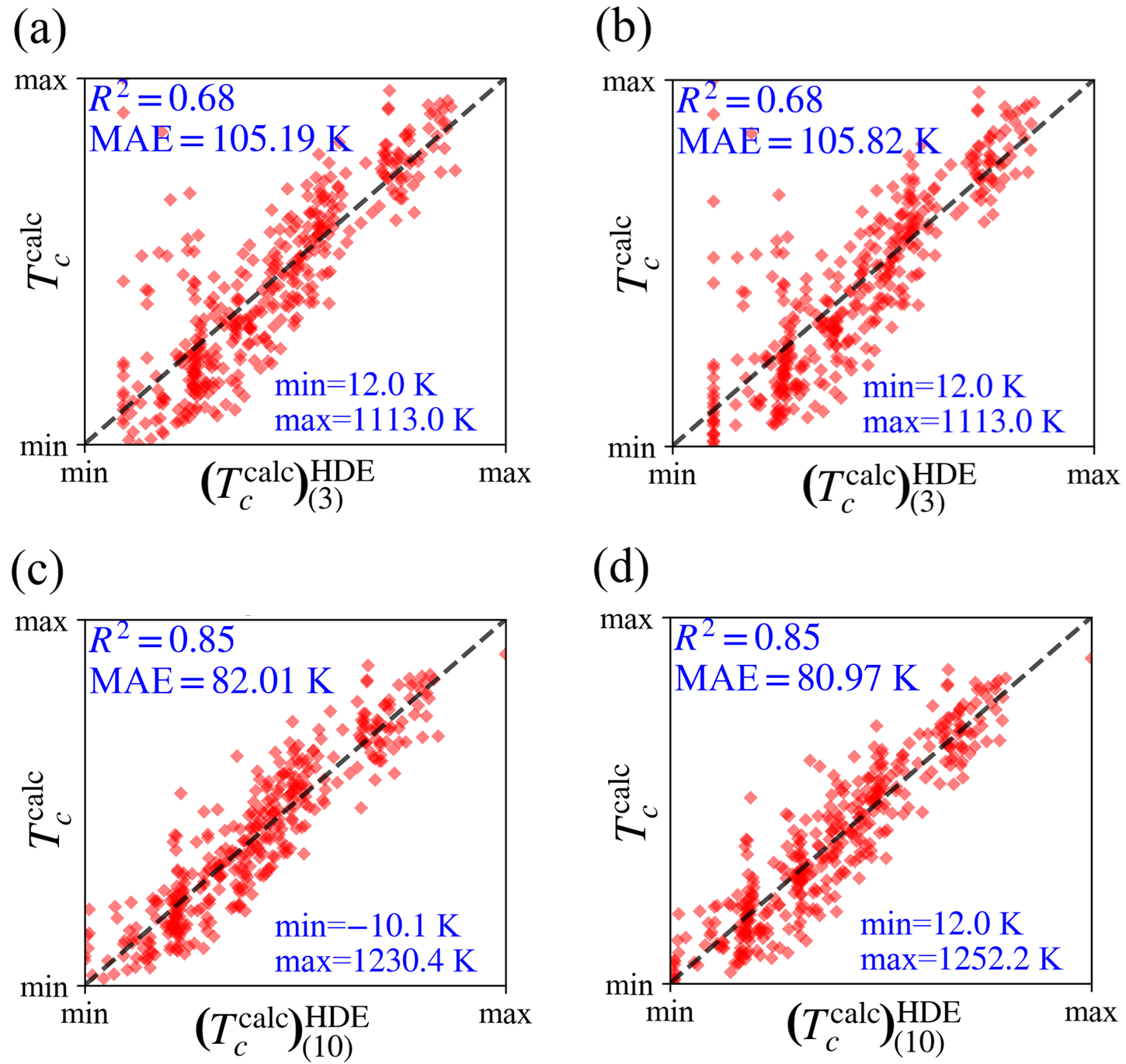}
    \caption{Results of the HDE runs where the target quantity is $T_c^{\rm calc}$, 
    the descriptors are the element-resolved magnetization amplitudes for the $3d$ 
    elements from Ti to Ni and their pairwise products, and the threshold $\mu_{\rm thr}$ on the magnetic moment amplitude is either 0 [left column, panels (a, c)] or $0.5 \mu_{\rm B}$ [right column, panels (b, d)]. We show the comparison between actual and HDE-predicted values of $T_c^{\rm calc}$ at generation $g=3$ [panels (a, b)] and at $g=10$ [panels (c, d)].  The $g=3$ expressions capture the dominant physical trends, while the $g=10$ expressions are the converged HDE predictions. The values of the coefficient of determination $R^2$ and the mean average error (MAE) are also given.}
    \label{fig:hde_TcMy}
\end{figure}

We now express the GGA values of $T_c^{\rm calc}$ in terms of the element-resolved 
magnetization amplitudes. The descriptor set ${\bf x}=\{M_{\rm Z},\, M_{\rm Z}M_{\rm Z'}\}$ 
contains the amplitudes $M_{\rm Z}$ defined in Eq.~\eqref{eq:my} and their pairwise 
products, where Z and Z' denote magnetic elements with open $3d$ shells (Ti--Ni). 
We find that $T_c^{\rm calc}$ depends mainly on the magnetization amplitudes of Fe, Co, 
and Mn, and increases with them. For $g=3$, the HDE expression can be written in the 
common form
\begin{align}
(T_c^{\rm calc})^{\rm HDE}_{(3)} &= a_0 + a_1\Big[ M_{\rm Fe}^{\alpha_1} 
+ a_2\, M_{\rm Co}^{\alpha_2} + a_3\, M_{\rm Mn}^{\alpha_3} \Big].
\label{eq:Tccalc_M_g3_common}
\end{align}
The fitted coefficients $\{a_i,\alpha_i\}$ for the two thresholds $\mu_{\rm thr}=0$ and 
$\mu_{\rm thr}=0.5\,\mu_{\rm B}$ are listed in Table~\ref{tab:Tccalc_M_coeffs}. The HDE 
expression is robust with respect to the threshold $\mu_{\rm thr}$ on the magnetic-moment 
amplitude. The agreement between the two thresholds is reflected by the similar coefficients 
in Table~\ref{tab:Tccalc_M_coeffs}, and is further illustrated in Fig.~\ref{fig:hde_TcMy}.
Note that the numerical values of $\{a_i,\alpha_i\}$ listed in Table~\ref{tab:Tccalc_M_coeffs} for the two thresholds are not directly physically comparable, since the coefficients have different physical dimensions. Indeed, $T_c^{\rm calc}$ is measured in K, whereas the magnetization amplitudes $M_{\rm Fe}$, $M_{\rm Co}$, and $M_{\rm Mn}$ are in $\mu_{\rm B}/\AA^3$. Consequently, $a_0$ always has units of K, while the units of $a_1$, $a_2$, and $a_3$ depend on the numerical values of $\alpha_1$, $\alpha_2$, and $\alpha_3$ \footnote{For the expressions of $T_c^{\rm exp}$ and $T_c^{\rm calc}$ as functions of the chemical proportions in Eq.~\eqref{eq:Tc_chemicalproportions}, the coefficients are physically comparable because the chemical proportions are dimensionless; in that case, the values of $\alpha_1$, $\alpha_2$, and $\alpha_3$ do not affect the dimensions of $a_1$, $a_2$, and $a_3$.}. Although the corresponding coefficients are therefore not physically identical between the two thresholds, their numerical values are close, so the predicted $T_c$ from Eq.~\eqref{eq:Tccalc_M_g3_common} is essentially unchanged. This is the relevant feature for predictive machine learning.

For $\mu_{\rm thr}=0$, we obtain $R^2_{(3)}=0.68$, compared with $R^2_{(3)}=0.53$ in 
Eq.~\eqref{eq:Tc_chemicalproportions}. Thus, $T_c^{\rm calc}$ is described more accurately in terms 
of the magnetization amplitudes than in terms of the chemical proportions. This result 
is consistent with Eq.~\eqref{eq:Tc_chemicalproportions}, since the magnetization amplitude of a given 
element generally increases with its chemical proportion. The robustness of the HDE 
expression is supported by the score analysis and the $4$-fold cross-validation. In 
the score analysis, the second-largest score values are $0.71$ at $g=1$, $0.88$ at $g=2$, 
and $0.94$ at $g=3$, all well below $1$ and therefore indicative of no close competition 
among the candidate descriptors. The $4$-fold cross-validation yields 
$R^2_{(3)\mathrm{CV}}=0.67$, close to $R^2_{(3)}=0.68$.

Finally, the HDE expression for $T_c^{\rm calc}$ in terms of the element-resolved 
magnetization amplitudes does not retain a finite offset $a_{(g)}$ in the limit $g\rightarrow\infty$. 
Although $a_{(3)}\simeq 110$ K is still finite, it becomes negligible at larger $g$. For instance, 
we find $a_{(10)}=-10$ K for $\mu_{\rm thr}=0$ and $a_{(10)}=16$ K for $\mu_{\rm thr}=0.5\,\mu_{\rm B}$. 
These values are small compared with the maximum value of $T_c^{\rm calc}\simeq 1300$ K, so the offset 
may be neglected and the HDE expression may be approximated as $(T_c^{\rm calc})^{\rm HDE}_{(g)}
\simeq b_{(g)} z^{\rm opt}_{(g)}$. This implies that $T_c^{\rm calc}$ vanishes when $z^{\rm opt}_{(g)}=0$, and hence when all $M_{\rm Z}$ vanish,  consistent with the expected physical 
behavior.

\begin{table}[t]
\caption{Numerical values of the coefficients $\{ a_i \}$ and exponents $\{ \alpha_i \}$ for the third-generation ($g=3$) HDE expression in Eq.~\eqref{eq:Tccalc_M_g3_common} for two values of the threshold $\mu_{\rm thr}$. We also list the values of the coefficient of determination $R^2$ at the third generation.}
\label{tab:Tccalc_M_coeffs}
\begin{ruledtabular}
\begin{tabular}{lcc}
 & $\mu_{\rm thr}=0$ & $\mu_{\rm thr}=0.5\,\mu_{\rm B}$ \\
\hline
$a_0$   & 107.8  & 110.8 \\
$a_1$   & 7768.0 & 5842.5 \\
$a_2$   & 0.41   & 0.41 \\
$a_3$   & 0.36   & 0.47 \\
$\alpha_1$ & 1.15 & 1.03 \\
$\alpha_2$ & 0.69 & 0.61 \\
$\alpha_3$ & 0.99 & 0.99 \\
$R^2$ & 0.68 & 0.68 \\
\end{tabular}
\end{ruledtabular}
\end{table}

\subsection{Emergent collective magnetic descriptor}
\label{sec:op}

In the previous section, we showed that the HDE expression of 
$T_c^{\rm calc}$ in terms of the element-resolved magnetization 
amplitudes has nearly negligible offset $a_{(g)}$ at sufficiently large $g$. 
This motivates the introduction of an effective collective magnetic descriptor $\phi^{\rm HDE}$ governing $T_c$. For HDE[$T_c^{\rm calc}$, $\{M_{\rm Z}, M_{\rm Z}M_{\rm Z'}\}$]  with $\mu_{\rm thr}=0.5\,\mu_{\rm B}$, the optimized HDE descriptor $\phi^{\rm HDE}\equiv z^{\rm opt}_{(5)}$ at fifth generation is
\begin{equation}
\begin{split}
\phi^{\rm HDE}
&=
\Big[
M_{\rm Fe}^{\alpha_{\rm Fe}}
+ A_{\rm Co}\, M_{\rm Co}^{\alpha_{\rm Co}}
+ A_{\rm Mn}\, M_{\rm Mn}^{\alpha_{\rm Mn}}
+ A_{\rm Cr}\, M_{\rm Cr}^{\alpha_{\rm Cr}}
\Big]\\
&\times \left(1+A_{\rm Ti}\, M_{\rm Ti}^{\alpha_{\rm Ti}}\right),
\label{eq:opg6}
\end{split}
\end{equation} 
and the values of the coefficients $\{A_i,\alpha_i\}$ are listed in Table~\ref{tab:phihde_coeffs}. The corresponding expression for the magnetic transition temperature is
\begin{equation}
(T_c^{\rm calc})^{\rm HDE}_{(5)}
=
37.96 + 1560.16\,\phi^{\rm HDE}.
\label{eq:Tc_phi}
\end{equation}
Since the offset is small compared with the overall scale of 
$T_c^{\rm calc}$, Eq.~\eqref{eq:Tc_phi} is approximately linear 
in $\phi^{\rm HDE}$. Unlike the analytical relations obtained from 
mean-field theory and RPA in Appendix~\ref{app:mfrpa}, 
$\phi^{\rm HDE}$ is determined numerically through the HDE 
procedure. Moreover, the HDE framework retains explicit information 
on the contributions from the individual magnetic elements.
\begin{table}[t]
\caption{Numerical values of the coefficients $\{ A_i \}$ and exponents $\{ \alpha_i \}$ for the effective collective magnetic descriptor $\phi^{\rm HDE}$ in Eq.~\eqref{eq:opg6}, which scales as the magnetic transition temperature $T_c$.}
\label{tab:phihde_coeffs}
\begin{ruledtabular}
\begin{tabular}{ccccccccc}
$\alpha_{\rm Fe}$ & $\alpha_{\rm Co}$ & $\alpha_{\rm Mn}$ & $\alpha_{\rm Cr}$ & $\alpha_{\rm Ti}$ & $A_{\rm Co}$ & $A_{\rm Mn}$ & $A_{\rm Cr}$ & $A_{\rm Ti}$\\
\hline
1.03 & 0.61 & 0.99 & 1.88 & 0.30 & 0.409 & 0.471 & 11.7 & 50.9\\
\end{tabular}
\end{ruledtabular}
\end{table}

\begin{figure}[!htb]
    \centering
    \includegraphics[width=1.0\linewidth]{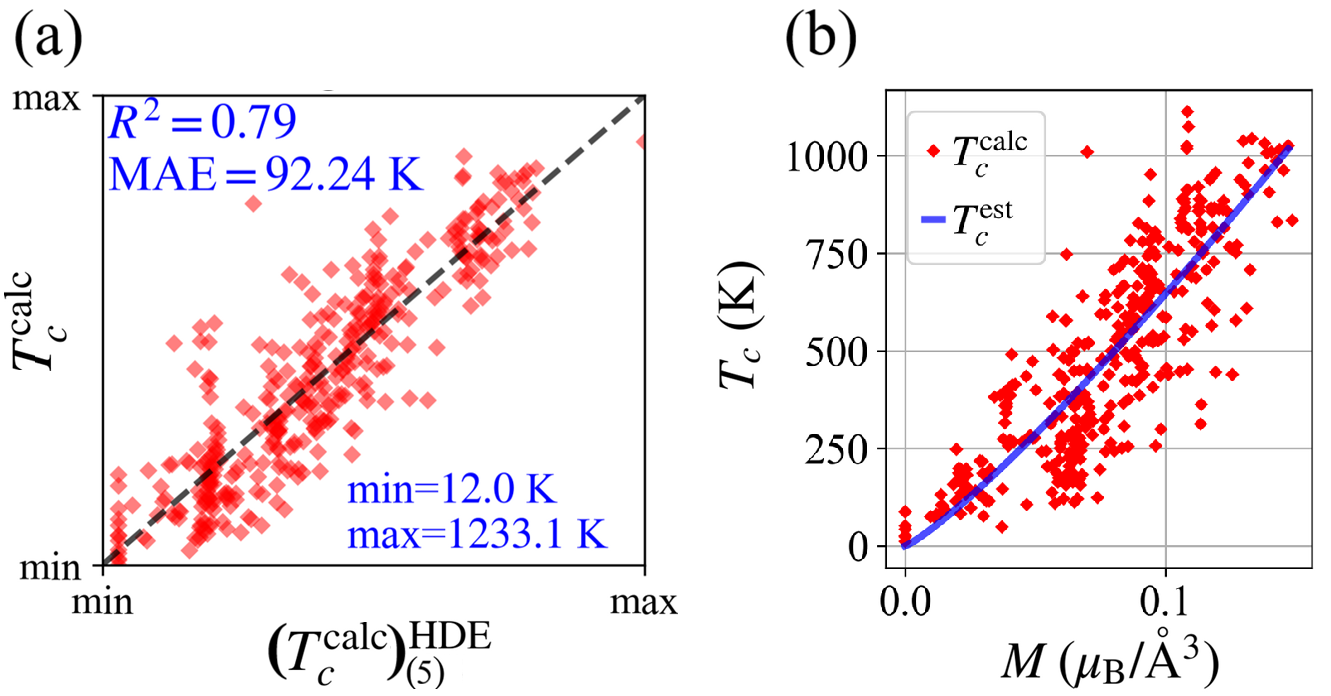}
    \caption{Comparison between the calculated magnetic transition temperature $T_c^{\rm calc}$ in the JuHemd database for Heusler compounds and the machine-learning expressions of $T_c^{\rm calc}$ obtained in this paper.
    (a): We compare $T_c^{\rm calc}$ with the fifth-generation HDE expression, which is $(T_c^{\rm calc})^{\rm HDE}_{(5)} \simeq 1560.16 ~\phi^{\rm HDE}$, and  $\phi^{\rm HDE}$ is the collective magnetic descriptor. The values of the coefficient of determination $R^2$ and the mean average error (MAE) are also given.
    (b): We show the dependence of $T_c^{\rm est} = 9681 \ M^{1.18}$ on the total magnetization amplitude $M$, which is obtained by fitting $T_c^{\rm calc}$ using a power law. The $M$ dependence of $T_c^{\rm calc}$ is also shown for comparison.}
    \label{fig:opg6}
\end{figure}

The coefficient of determination at fifth generation is $R^2_{(5)}=0.79$, indicating that $\phi^{\rm HDE}$ captures the main trend of $T_c$. The dependence of $T_c^{\rm calc}$ on $(T_c^{\rm calc})^{\rm HDE}_{(5)}\simeq 1560.16\,\phi^{\rm HDE}$ is depicted in \dum{Fig.~\ref{fig:opg6}(a)}. As expected, $T_c^{\rm calc}$ increases with $\phi^{\rm HDE}$ and vanishes when $\phi^{\rm HDE}=0$, which also occurs when all $M_{\rm Z}$ vanish since $\phi^{\rm HDE}$ is built from the magnetization amplitudes of Fe, Co, Mn, Cr, and Ti. The descriptor $\phi^{\rm HDE}$ is also approximately homogeneous to magnetization: The exponent $1.03$ in $M_{\rm Fe}^{1.03}$ is close to unity, and $z^{\rm opt}_{(g+1)}$ has the same dimension as $z^{\rm opt}_{(g)}$ by construction, as seen in Eq.~\eqref{eq:xg}. Overall, the HDE procedure captures a simple and physically transparent collective mechanism, with $b_{(5)}=1560.16$ K/($\mu_{\rm B}/\AA^3$) acting as an effective scale factor for this Heusler family.

As a simpler benchmark, we test whether $T_c$ can be described directly in terms of the total magnetization amplitude. This provides a reference for assessing how much of the magnetic transition temperature trend is already contained in a single collective variable, we therefore consider
\begin{equation}
M = \sum_{\rm Z} M_{\rm Z}.
\label{eq:op0}
\end{equation}
The previous sum runs over $3d$ magnetic elements Z from Ti to Ni. Using the threshold $\mu_{\rm thr}=0.5\,\mu_{\rm B}$ in Eq.~\eqref{eq:my}, we first compute the element-resolved amplitudes $M_{\rm Z}$ and then construct the total magnetization amplitude $M$ from Eq.~\eqref{eq:op0}. A power-law fit of the form $T_c^{\rm calc} = b \,M^\alpha$ yields the estimate
\begin{equation}
T_c^{\rm est} = 9681\, M^{1.18},
\label{eq:TcestM}
\end{equation}
\dum{which is represented in Fig.~\ref{fig:opg6}(b).} 
The resulting fit gives $R^2=0.71$, which is lower than the value \dum{0.79} obtained for the HDE collective descriptor in Eq.~\eqref{eq:opg6}. The exponent $\alpha=1.18$ is also robust, changing only to $1.29$ when setting $\mu_{\rm thr}=0\,\mu_{\rm B}$. 
Furthermore, the value of $\alpha$ is close to $1$, so that $M^{\alpha}$ is almost homogeneous to $M$, and the fitted coefficient $b$ is almost homogeneous to $b_{(5)}$ in the above-discussed HDE expression. Overall, the total magnetization captures the dominant magnetic scale governing $T_c$, while the HDE descriptor provides a more accurate description by retaining element-specific contributions and nonlinear couplings.

\subsection{Dependence of element-resolved magnetization amplitudes on chemical proportions}
\label{sec:results4}
In the previous sections, we used HDE to examine the dependence of $T_c$ on the chemical proportions, $d_{\rm Y}$. For the calculated transition temperatures, $T_c^{\rm calc}$, we further expressed the results in terms of the element-resolved magnetization amplitudes, $M_{\rm Z}$. We now use the HDE approach to analyze the relation between these two descriptor sets. In general, this relation is nontrivial, since it depends on several factors, including the local geometry (e.g., bond lengths and bond angles), the electronic configuration of the magnetic ions, the crystal field, and the exchange interactions between neighboring magnetic moments. Here, we adopt a simplified approach in which the chemical proportions $d_{\rm Y}$ are used as descriptors and the magnetization amplitudes $M_{\rm Z}$ as targets. The descriptor set is ${\bf x}=\{d_{\rm Y},\, d_{\rm Z}d_{\rm Y}\}$, where Z denotes magnetic elements with open $3d$ shells (Ti--Ni), and Y denotes any chemical element. To exclude possible induced magnetic moments that would otherwise complicate the analysis, we apply the threshold $\mu_{\rm thr}=0.5\,\mu_{\rm B}$.

We examine the HDE expressions for $M_{\rm Fe}$ and $M_{\rm Co}$ at $g=2$, and for $M_{\rm Mn}$ at $g=1$. At these values of $g$, the coefficient of determination $R^2$ is already high (Table~\ref{tab:Mz_hde_coeffs}), indicating that the dominant dependence of $M_{\rm Fe}$, $M_{\rm Co}$, and $M_{\rm Mn}$ is captured without requiring higher-order expansions.
For clarity, the HDE expressions can be written in the compact forms
\begin{align}
(M_{\rm Fe})^{\rm HDE}_{(2)} &= A_{\rm Fe}\, d_{\rm Fe}^{\alpha_{\rm Fe}}
\left[1 - B_{\rm Fe}\left(d_{\rm Fe}d_{\rm Mn}\right)^{\beta_{\rm Fe}}\right], \label{eq:MFe} \\
(M_{\rm Co})^{\rm HDE}_{(2)} &= A_{\rm Co}\, d_{\rm Co}^{\alpha_{\rm Co}}
\left[1 + B_{\rm Co}\left(d_{\rm Co}d_{\rm Fe}\right)^{\beta_{\rm Co}}\right], \label{eq:MCo} \\
(M_{\rm Mn})^{\rm HDE}_{(1)} &= A_{\rm Mn} d_{\rm Mn}^{\alpha_{\rm Mn}}.
\end{align}
The fitted coefficients are listed in Table~\ref{tab:Mz_hde_coeffs}. The offsets are negligible in all cases, with $|a_{(g)}|<0.021\,\mu_{\rm B}/\mathrm{\AA}^3$,
and are therefore omitted from the formula above. The relatively large $R^2$ values already obtained at low generation ($g\leq2$), together with the close agreement between the $R^2$ and $4$-fold cross-validation scores $(R^2_\text{CV})$ in Table~\ref{tab:Mz_hde_coeffs}, indicate that the HDE expressions are robust. At $g=1$, $M_{\rm Fe}$, $M_{\rm Co}$ and $M_{\rm Mn}$ depend primarily on $d_{\rm Fe}$, $d_{\rm Co}$ and $d_{\rm Mn}$, respectively, consistent with the intuitive expectation that the local magnetization amplitude of a magnetic element is governed mainly by its own chemical proportion.
\begin{table}[t]
\caption{Numerical values of the coefficients $\{ A_i, B_i \}$ and exponents $\{ \alpha_i, \beta_i \}$ for the HDE expressions of the magnetization amplitudes in terms of the chemical proportions. The numerical values of the offsets are negligible and are
not shown explicitly. We also list the values of the
coefficient of determination $R^2$ at the generation $g$.}
\label{tab:Mz_hde_coeffs}
\begin{ruledtabular}
\begin{tabular}{lcccccc}
Target & $g$ & $A$ & $\alpha$ & $B$ & $\beta$ & $R^2$ ($R^2_{\rm CV}$) \\
\hline
$M_{\rm Fe}$ & 2 & 0.192 & 1.24 & 8.21 & 0.99 & 0.87 (0.86) \\
$M_{\rm Co}$ & 2 & 0.094 & 1.24 & 3.62 & 0.99 & 0.90 (0.89) \\
$M_{\rm Mn}$ & 1 & 0.141 & 0.42 & --- & --- & 0.91 (0.90) \\
\end{tabular}
\end{ruledtabular}
\end{table}

\begin{figure}[!htb]
    \centering    
    \includegraphics[width=1.0\linewidth]{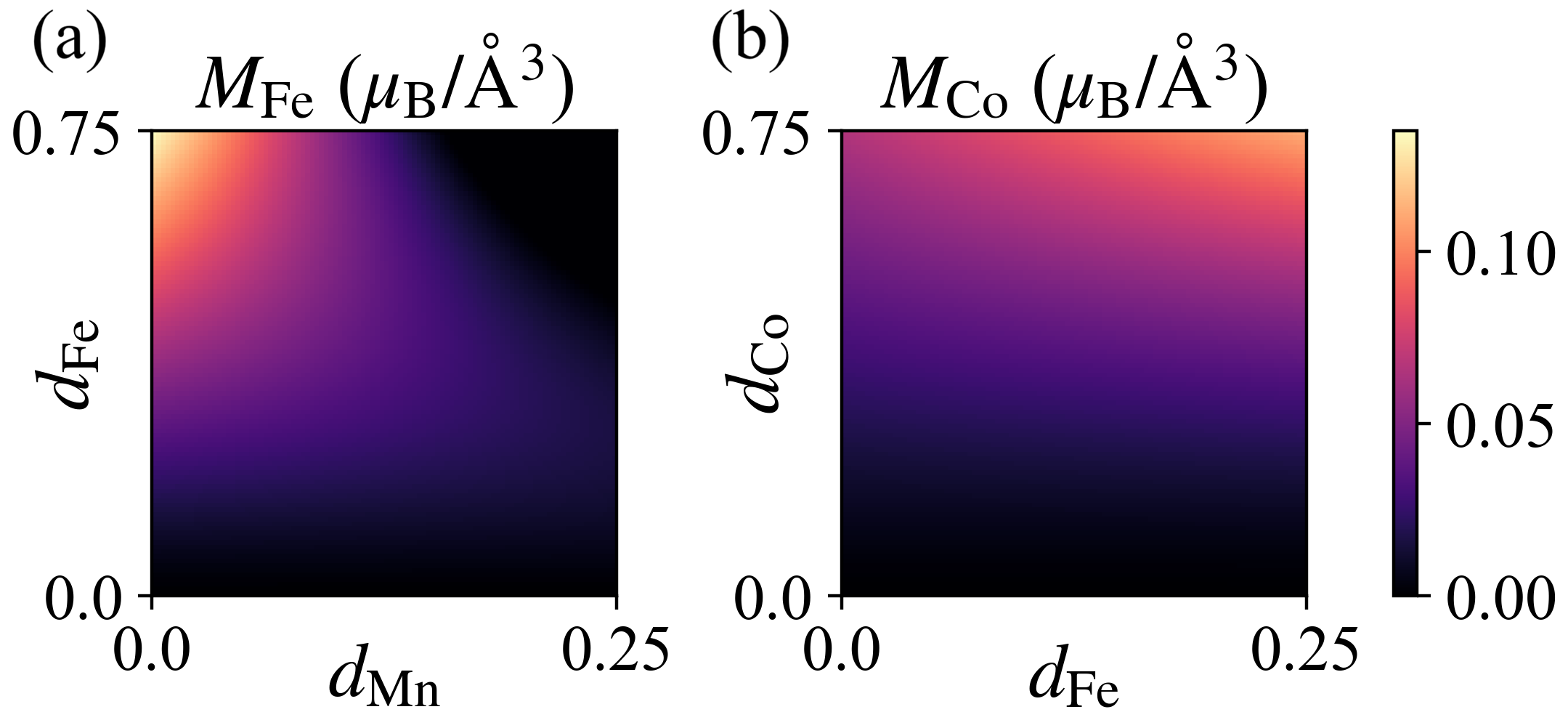}
    \caption{Representations of the HDE expressions of the magnetization amplitudes for Fe [panel (a), corresponding to Eq.~\eqref{eq:MFe}] and Co [panel (b), corresponding  to Eq.~\eqref{eq:MCo}], in terms of the chemical proportions $d_{\rm Fe}$, $d_{\rm Co}$, and $d_{\rm Mn}$ of Fe, Co, and Mn. The magnetization amplitudes are in $\mu_{\rm B}/\AA^3$ while the chemical proportions are dimensionless. The representations illustrate the increases in $M_{\rm Fe}$ and $M_{\rm Co}$ with $d_{\rm Fe}$ and $d_{\rm Co}$, respectively.}
    \label{fig:hde_Mydy}
\end{figure}

At $g=2$, $M_{\rm Fe}$ is reduced by increasing $d_{\rm Mn}$, whereas $M_{\rm Co}$ is enhanced by increasing $d_{\rm Fe}$. These trends are physically intuitive and can be understood from the magnetic couplings between the constituent elements. In particular, Fe and Co moments are ferromagnetically coupled in several Heusler compounds, such as Fe$_2$CoGa and Fe$_2$CoZn~\cite{Dannenberg2010}, so that the presence of Fe enhances the Co magnetic moment. In contrast, Mn moments tend to couple antiferromagnetically in Mn-containing Heusler alloys, for example in Fe$_{3-x}$Mn$_x$Ga~\cite{Wang2023}, which leads to a reduction of the Fe magnetic moment in the presence of Mn. The extracted HDE dependencies of $M_{\rm Fe}$ and $M_{\rm Co}$ are illustrated in \dum{Fig.~\ref{fig:hde_Mydy}}.

\section{Discussion and conclusion}
\label{sec:disc}

\subsection{HDE as an interpretable approach for $T_{c}$}
Our HDE result for collinear Heusler magnets is consistent with previous studies on ferromagnets and Heusler compounds. In Refs.~\cite{Belot2023,Jung2024,Hilgers2025},  $T_c$ was found to increase with the presence of Fe and/or Co in ferromagnets; HDE reproduces the same qualitative trend and extends it to ferrimagnets. In addition,  Ref.~\cite{Hilgers2025} reported a correlation between $T_c$ and the sum\footnote{In Ref.~\cite{Hilgers2025}, the sum is taken over site indices, whereas here it is taken over chemical elements. We use the latter definition because element-resolved quantities are more directly linked to chemical composition and therefore provide a more natural choice of variables that can be adjusted to control $T_c$ in experimental synthesis of Heusler compounds.} of magnetic-moment amplitudes on each site in the unit cell shown in Fig.~\ref{fig:cryst}. This is consistent with our HDE expressions of $T_c$ in terms of the magnetization amplitudes.

The accuracy of HDE is also comparable to that of other machine-learning approaches. In our calculations, the coefficient of determination converges to  $R^2_{\infty}\gtrsim 0.85$ when $g$ is increased beyond $3$, which is comparable to  the range $R^2\simeq 0.53$--$0.93$ reported in Refs.~\cite{Nelson2019,Belot2023,Hilgers2025,Jung2024}. This accuracy is 
limited by  the simple power-law form assumed for $z_{(g)}$ in Eqs.~\eqref{eq:x1} and~\eqref{eq:xg}. More flexible forms could further 
improve the accuracy of HDE, at the cost of increased computational effort, and will be considered in future work.

\subsection{Role of Fe and Co in magnetic exchange}
The increase in $T_c$ with increasing Fe and Co content observed in this work and in Refs.~\cite{Belot2023,Jung2024,Hilgers2025} is physically consistent. While Mn has a large atomic Hund's-rule moment, elemental Mn does not form a robust ferromagnet but instead exhibits complex antiferromagnetic ordering with a relatively low characteristic transition scale ($T \sim 95$~K in $\alpha$-Mn). By contrast, Fe and Co are prototypical ferromagnets with high Curie temperatures in their elemental phases ($T_c^{\mathrm{Fe}}\approx 1043$~K and $T_c^{\mathrm{Co}}\approx 1400$~K), reflecting strong ferromagnetic exchange. This behavior is consistent with the fact that $T_c$ in bulk materials is mainly determined by exchange interactions~\cite{bruno1991theory}. In mean-field theory, one has
\begin{equation}
    T_c \sim J_\text{max}(q)/3k_{\rm B},
    \label{eq:tcmfJ}
\end{equation}
where $q$ is a wavevector, $J(q)$ is the exchange interaction in Fourier space~\cite{Bouaziz2022}, and $k_{\rm B}$ is the Boltzmann constant.
See Appendix~\ref{app:mfrpa} for a detailed derivation of Eq.~\eqref{eq:tcmfJ}.
Although explicit $J_{ij}$ values are not available in the JuHemd database, Eq.~\eqref{eq:tcmfJ} suggests that the dominant exchange couplings are primarily governed by the Fe and Co content. This is consistent with their role in stabilizing ferromagnetic interactions in these compounds. Lastly, the Heisenberg Hamiltonian used in the Monte Carlo simulations assumes rigid spin moments, i.e., it neglects longitudinal fluctuations. Despite this approximation, the HDE analysis consistently reproduces the same Fe- and Co-driven trends in both the calculated and experimentally determined $T_c$ as shown in Eq.~\eqref{eq:Tccalc_M_g3_common}. This agreement indicates that the dominant behavior is well captured within an effective fixed-moment description.

\subsection{Conclusion}
We have used hierarchical dependence extraction (HDE) to analyze how the magnetic transition temperature $T_c$ depends on the chemical composition in collinear Heusler magnets. Applied to both the experimental and GGA-based datasets, HDE identifies the same dominant chemical trends and shows that higher $T_c$ is mainly associated with higher proportion of Fe, Co, and Mn in the chemical composition. The resulting HDE expressions are compact and interpretable: At low generation, they reproduce the main trends of $T_c$ with chemical proportions, and when element-resolved magnetization amplitudes are used as intermediate descriptors, the description becomes more accurate. The HDE results are also robust under score analysis and cross-validation, indicating that the extracted descriptor structure is stable.

The analysis further shows that the total magnetization provides a useful coarse-grained benchmark, but that the HDE descriptor retains additional element-specific information and nonlinear couplings that improve the description of $T_c$. In this sense, the HDE framework captures a physically meaningful hierarchy from chemical composition to local magnetization amplitudes and then to the magnetic transition temperature. Overall, the present work shows that HDE can extract simple and physically transparent scaling relations from Heusler datasets and may be useful for analyzing other functional magnetic materials.

\section*{Acknowledgments}
We thank Johannes Wasmer for useful discussions. J.-B. M. acknowledges funding from the Special Postdoctoral Researcher Program at RIKEN. J. B. was supported by the Alexander von Humboldt Foundation through the Feodor Lynen Research Fellowship for Postdocs. This work was supported by the RIKEN TRIP initiative (RIKEN Quantum, Advanced General Intelligence for Science Program, Many-body Electron Systems). R. A. acknowledges the financial support by Grant-in-Aids for Scientific Research (JSPS KAKENHI) Grant No. 25H01252.

\appendix

\section{Appendix: Previous machine learning works on the prediction of $T_c$ in ferromagnets and Heusler compounds}
\label{app:mlworks}

\headline{Previous works in the literature use various machine learning methods to predict $T_c$ in ferromagnets and Heusler magnets.}
We summarize the key results in Table~\ref{tab:litt}.

\begin{table*}[!htb]
\centering
\caption{Summary of previous machine learning analyses of $T_c$ in ferromagnets (FM) and Heusler magnets. We summarize the categories of coumpounds that are treated, the number $N_{\rm ent}$ of entries in the dataset, the methods that are employed, the obtained values of the coefficient of determination $R^2$, and the main descriptors that control $T_c$ when available.
The methods employed in these works include neural networks (NN)~\cite{McCulloch1943, Rosenblatt1958, Rumelhart1986, LeCun1998, Krizhevsky2012, Goodfellow2016},
nearest-neighbor pattern classification (kNN)~\cite{Cover1967},
random forests (RF)~\cite{Breiman2001},
Ridge regression (RR)~\cite{Hoerl1970},
kernel Ridge regression (KRR)~\cite{Saunders1998},
gradient boosting regression (GBR)~\cite{Friedman2001greedy, Friedman2000additive, Chen2016xgboost},
and gradient-boosted statistical feature selection (GBFS)~\cite{Jung2024} combined with the SHapley Additive exPlanations (SHAP) framework~\cite{Shapley1953value}.
Our HDE result on ferromagnets (FM) and ferrimagnets (FiM) is also shown for comparison.
}
\begin{ruledtabular}
\begin{tabular}{lllllll}
Ref. & Nelson \textit{et al.}~\cite{Nelson2019} & Belot \textit{et al.}~\cite{Belot2023} & Hilgers \textit{et al.}~\cite{Hilgers2025} & Jung \textit{et al.}~\cite{Jung2024} & Court \textit{et al.}~\cite{Court2021} & This work\\
Compounds & FM & FM & Heusler & FM & Heusler (FM) & Heusler (FM +FiM) \\
$N_{\rm ent}$ & $\simeq$ 2500 & 2557 (dataset 1) & 408 & 35000 (2 datasets) & 4848 & 653 (experiments) \\
 &  & 3194 (dataset 2) &  & & & 372 (calculations)\\
Methods & RR, NN, KRR, RF & RF, kNN & RF, Extra Trees\footnote{We show only the regression scores obtained using ensemble models.} & GBFS, SHAP & GBR & HDE\\
$R^2$ & 0.53-0.81 & & $\lesssim 0.85$ & 0.93 & 0.71 & $\lesssim 0.91$\\
Main descriptors & --- & $d_{\rm Fe}$, $d_{\rm Co}$ & $\sum_{i}^{\rm site} |\mu_{i}|$ & $\tilde{\mu}$, $d_{\rm Co}$, $N_d$, $V_{\rm cell}$ & --- & $d_{\rm Z}$ and $M_{\rm Z}$ (Z = Fe, Co, Mn)\\
\end{tabular}
\end{ruledtabular}
\label{tab:litt}
\end{table*}

\section{Appendix: Formula for $T_c$ in mean-field and random phase approximation theories}
\label{app:mfrpa}

Here, we propose different approaches to understand the dependence of $T_c$ on the magnetic moment length, where $T_c$ is the transition temperature from a magnetically ordered phase (FM, FiM, or antiferromagnetic) to a paramagnetic phase.
The idea is to examine the dependence of the magnetic exchange interactions on the moments' lengths 
and then derive an expression of $T_c$ as a function of the magnetic moments using a mean-field approximation or the RPA.
These expressions can be compared to the HDE expression: The deviations can be understood as higher order corrections coming from the DFT self-consistent treatment of the spin splitting and the classical Monte-Carlo approach. (The latter goes beyond the mean-field approach and includes nonlocal correlations.)

\subsection{General approach}
We follow the methodology in Ref.~\cite{Rosengaard1997} to 
incorporate the contribution of longitudinal spin fluctuations into the magnetic interactions. This was inspired from the Stoner-Wohlfarth model, which is a Landau-Ginzburg-like theory where
the energy expansion in the magnetization $M$ reads:
\begin{equation}
    E_{\rm FM} = E_{\rm PM} + \sum_{i=1}^{n}A^{\rm FM}_iM^{2i},
\end{equation}
where $E_{\rm FM}$ is the energy of the ferromagnetic state, $E_{\rm PM}$ is the energy of the paramagnetic state,
and $A^{\rm FM}_i$ are the expansion coefficients.
This form is valid near the phase transition.
It includes contributions from both local and nonlocal exchange-correlation ($B_{xc}$ at the DFT level).
It neglects the vector character of the magnetic moments.
Nevertheless, it is possible to model in DFT the magnetic moment and the formation energy of a constrained moment, either by scaling $B_{xc}$ or adding a large longitudinal magnetic field.
Note that only terms that are even in the magnetization will contribute, since the energy
must remain invariant under time-reversal symmetry ($M \rightarrow -M$). The parameters
$A^{\rm FM}_i$ can be obtained from constrained DFT FM calculations~\cite{Moruzzi1986}. 

\subsection{Transverse fluctuations}
\headline{To incorporate the vector character of the magnetization, and assuming that the magnetic moments are robust and localized,
we can describe the transverse
energy fluctuations using the Heisenberg Hamiltonian}:
\begin{equation}
H = -\frac{1}{2}\sum_{ij} J_{ij}\,\vec{e}_{i}\cdot\vec{e}_{j},
\end{equation}
where $\vec{e}_{i}$ is the direction of the magnetic moment $\vec{m}_{i} = m_i \vec{e}_{i}$, and the $J_{ij}$ are the isotropic exchange interactions.
The $J_{ij}$ contain the information on the moment's length and are responsible for the magnetic transition near $T_c$.
Other higher order interactions are present; 
although they can impact the magnetic order at temperature lower than $T_c$, they do not play a major role in determining $T_c$.

The $J_{ij}$ can be obtained using the infinitesimal rotation method starting from a FM ordered state~\cite{Szilva2023}:
\begin{equation}
  J_{ij} = \frac{2}{\pi}\int dE\, \text{Tr}_{L} B_{i}\,G^{\uparrow}_{ij}\,B_j\,G^{\downarrow}_{ji},
  \label{lkag_spin}
\end{equation}
where $B_i$ is the exchange spin splitting,
and Tr$_{L}$ is the trace over the orbital angular momentum $L=(l,m)$.
(We have $B_{i}=B^{LL^\prime}_{i}$ and $G^{\sigma}_{ij} = G^{\sigma,LL^\prime}_{ij}$, where $\sigma = \uparrow, \downarrow$ is the spin.)
\\

We connect $J_{ij}$ in Eq.~\eqref{lkag_spin} to the magnetic moments $m_{i}$ in the framework of mean-field theory. If we use the Hubbard model approach
$B_{i} = Um_{i}/2$ (where $U_i$ is the Hubbard parameter) or the Stoner model approach~\cite{Rosengaard1997} $B_{i} = m_i I_i$ (where $I_i$ is the Stoner parameter), Eq.~\eqref{lkag_spin} yields $J_{ij}\propto m_{i} m_{j}$. However, $G_{ij}$ also has a dependence on the spin splitting because we 
have $G_{ij} = {G}^{0}_{ij}\,\sigma_{0} + \vec{G}_{ij}\cdot\vec{\sigma}$,
where $\sigma_{0}$ is the identity and $\vec{\sigma}$ is the vector of the Pauli matrices. This makes the assumption $J_{ij}\propto m_{i} m_{j}$ too simplified, since $J_{ij}$ will acquire a nonlinear dependence.
To go beyond the above assumption, the generalization of the Stoner-Wohlfarth approach in a constrained ferromagnetic state leads to an expansion of the $J_{ij}$ in power series of the $m_i m_j$, as~\cite{Rosengaard1997}:
\begin{equation}
  J_{ij} = \sum_{k=1}^{k_\text{max}} J^{k}_{ij}\,\langle{m_im_j}\rangle^{2k},
  \label{eq:jijmimj}
\end{equation}
where $J_{ij}^{k}$ are the expansion coefficients. (Note that $k$ starts 
from one, so that $J_{ij}$ vanishes when the magnetic moments vanish). 
In practice, it is generally sufficient for conventional ferromagnets 
(Fe, Co, Ni) to consider a cutoff of $k_\text{max}=3$.

We assume the average of the magnetic moments, $\langle m_im_j \rangle$ is taken as the simple average of the local moments, that is,
\begin{equation}
    \langle m_im_j \rangle=\frac{m_i+m_j}{2}.
    \label{eq:assu}
\end{equation}
This assumption has a drawback, that even when $m_i=0$ or $m_j=0$, the $J_{ij}$ does not vanish. But this is not an issue since at non-zero temperature or in non-collinear constrained states, the local moments do 
not deviate sufficiently from the ferromagnetic magnetization value.
Using Eq.~\eqref{eq:assu}, we can rewrite Eq.~\eqref{eq:jijmimj} as
\begin{equation}
  J_{ij} = \sum_{k=1}^{k_\text{max}} \frac{J^{k}_{ij}}{2^{2k}}\left(m_i+m_j\right)^{2k},
  \label{eq:expj1}
\end{equation}
and we can expand the power series using the binomial theorem
\begin{equation}
 \left(A+B\right)^n = \sum_{m=0}^{n} \frac{n!}{m!(n-m)!} A^{n-m}B^{m},   
\end{equation}
so that
\begin{eqnarray}
\left(m_i+m_j\right)^{2k} &= \sum_{n=0}^{2k} \frac{2k!}{n!(2k-n)!} m^{2k-n}_i\,m^{n}_j,\\
& =  \sum_{n=0}^{2k} F(n,2k)\,m^{2k-n}_i\,m^{n}_j. 
\label{eq:expj2}
\end{eqnarray}
Using Eqs.~\eqref{eq:expj1} and~\eqref{eq:expj2}, we deduce the expansion of $J_{ij}$ as:
\begin{eqnarray}
J_{ij} &=& \sum_{k=1}^{k_\text{max}} \sum_{n=0}^{2k} \frac{J^{k}_{ij}}{2^{2k}} F(n,2k)\,m^{2k-n}_i\,m^{n}_j,\label{eq:Jij}
\end{eqnarray}
which is the general connection between the magnetic interactions and the length of the magnetic moments in the Stoner-Wohlfarth-like approach.

\subsection{Transition temperature}
Now, we express $T_c$ using Eq.~\eqref{eq:Jij}. We consider the simple case of a ferromagnetic system with a single atom per unit cell, and the Heisenberg Hamiltonian
\begin{equation}
H = -\frac{1}{2}\sum_{ij} J_{ij}\,\vec{e}_{i}\cdot\vec{e}_{j}. 
\end{equation}
In this case, the transition temperature in mean field limit (MF) is 
simply given by~\cite{Bouaziz2022}:
\begin{equation}
 T_{c}^{\rm MF} = \frac{J(\vec{q}_\text{max})}{3k_\text{B}},
\end{equation}
where
\begin{equation}
  J(\vec{q}) = \sum_{j}e^{i\vec{q}\cdot(\vec{R}_{j}-\vec{R}_{0})}J_{0j}
\end{equation}
is the Fourier transform of the magnetic interactions, $k_\text{B}$ is the
Boltzmann constant, and $\vec{q}_\text{max}$ is the value of $\vec{q}$ that maximizes $J(\vec{q})$. Using Eq.~\eqref{eq:Jij}, we obtain:
\begin{eqnarray}
  J(\vec{q}) &=& \sum_{j}e^{i\vec{q}\cdot(\vec{R}_{j}-\vec{R}_{0})}J_{0j},\\
  &= &\sum_{j}e^{i\vec{q}\cdot(\vec{R}_{j}-\vec{R}_{0})}\sum_{k=1}^{k_\text{max}} \sum_{n=0}^{2k} \frac{J^{k}_{0j}}{2^{2k}} F(n,2k)\,m^{2k-n}_0\,m^{n}_j,\\
&=& \sum_{k=1}^{k_\text{max}}\sum_{n=0}^{2k}\,F(n,2k)\,m^{2k-n}_0\,m^{n}_j
\sum_{j}e^{i\vec{q}\cdot(\vec{R}_{j}-\vec{R}_{0})} \frac{J^{k}_{0j}}{2^{2k}},\\
&=& \sum_{k=1}^{k_\text{max}}\sum_{n=0}^{2k}\,F(n,2k) \mathcal{G}(k,\vec{q})\,m^{2k-n}_0\,m^{n}_j, \label{eq:jq}
\end{eqnarray}
where the function
\begin{equation}
\mathcal{G}(k,\vec{q}) = \sum_{j}
\frac{J^{k}_{0j}}{2^{2k}}e^{i\vec{q}\cdot(\vec{R}_{j}-\vec{R}_{0})}
\end{equation}
absorbs the geometrical complexity due to the $\vec{q}$-dependence (or real space dependence).
In our simplified case, $m_j=m_0$ because we consider a ferromagnetic system with a single atom per unit cell, and $T_c$ further simplifies as:
\begin{equation}
\begin{split}
T_{c}^{\rm MF} &=
\frac{1}{3k_\text{B}}\sum_{k=1}^{k_\text{max}}\sum_{n=0}^{2k}\,F(n,2k)\mathcal{G}(k,\vec{q}_\text{max})\,m^{2k}_0,
\label{eq:tcmf}
\end{split}
\end{equation}
so that, at the mean field level, $T_c$ can be expanded in even powers of $m_0$.
\\
Next, we consider the RPA in the same simple case of a ferromagnet with atom per unit cell. Here, $T_c$ is given by~\cite{Turek2006}:
\begin{equation}
T^{\rm RPA}_{c} = \frac{1}{3k_\text{B}}\left(\frac{1}{N_{\vec{q}}}\sum_{\vec{q}}\frac{1}{J(0)-J(\vec{q})}\right)^{-1},
\label{eq:tcrpa0}
\end{equation}
where $N_{\vec{q}}$ is the number of wave vectors. Using Eq.~\eqref{eq:jq} with $m_j=m_0$, we obtain:
\begin{equation}
J(0)-J(\vec{q}) =  \sum_{k=1}^{k_\text{max}}\sum_{n=0}^{2k}\,F(n,2k)\left[\mathcal{G}(k,0)-\mathcal{G}(k,\vec{q})\right]\,m^{2k}_0,
\end{equation}
and we deduce the analytical expression of $T_c$ in the RPA:
\begin{widetext}
\begin{equation}
T^{\rm RPA}_{c} = \frac{1}{3k_\text{B}}\left(\frac{1}{N_{\vec{q}}}\sum_{\vec{q}}\frac{1}{\sum_{k=1}^{k_\text{max}}\sum_{n=0}^{2k}\,F(n,2k)[\mathcal{G}(k,0)-\mathcal{G}(k,\vec{q})]\,m^{2k}_0}\right)^{-1}. 
\label{eq:tcrpa}
\end{equation}    
\end{widetext}
In both the mean-field and RPA expressions of the transition temperature in Eqs.~\eqref{eq:tcmf} and~\eqref{eq:tcrpa}, the main insights are (i) the ordering temperature scales with the magnetic moment amplitude $m_0$, and (ii) the ordering temperature vanishes when $m_0$ vanishes. In the mean-field approximation, this dependence can be explicitly expanded in even powers of $m_0$, whereas in RPA it enters through a more complex nonlinear momentum-dependent expression. For completeness, note that more sophisticated methods such as Monte-Carlo simulations include additional effects, e.g. short-range correlations, leading to even more complex expressions for the ordering temperature compared to Eqs.~\eqref{eq:tcmf} and~\eqref{eq:tcrpa}. Nevertheless, the qualitative dependence on the magnetic moment amplitude remains consistent, as evidenced by the HDE relation between $T^\text{calc}_c$ obtained from Monte-Carlo results and $m_0$.
\bibliographystyle{apsrev4-2}

%

\end{document}